\documentclass[12pt,reqno]{amsart}
\allowdisplaybreaks[4]
\usepackage{graphicx} 
\usepackage{amssymb}
\usepackage{amsmath}
\usepackage{cite}

\newcommand{\be}{\begin{equation}}
\newcommand{\ee}{\end{equation}}
\newcommand{\bea}{\begin{eqnarray}}
\newcommand{\eea}{\end{eqnarray}}
\newcommand{\ba}{\begin{array}}
\newcommand{\ea}{\end{array}}
\newcommand{\bean}{\begin{eqnarray*}}
\newcommand{\eean}{\end{eqnarray*}}

\newcommand{\pa}{\partial}

\begin{document}
\title[Darboux]{ The Rogue Wave and breather solution  of the Gerdjikov-Ivanov equation}
\author{Shuwei Xu\dag,Jingsong He$^*$ \dag}

\thanks{$^*$ Corresponding author: hejingsong@nbu.edu.cn,jshe@ustc.edu.cn}

 \maketitle
\dedicatory {  \dag \ Department of Mathematics, Ningbo University,
Ningbo , Zhejiang 315211, P.\ R.\ China\\}

\begin{abstract}
The Gerdjikov-Ivanov (GI) system of $q$ and $r$ is defined by a
quadratic polynomial spectral problem with $2 \times 2$ matrix
coefficients. Each element of the matrix of n-fold Darboux
transformation of this system is expressed by a ratio of
$(n+1)\times (n+1)$ determinant and  $n\times n$ determinant of
eigenfunctions, which implies the determinant representation of
$q^{[n]}$ and $r^{[n]}$ generated from known solution $q$ and $r$.
By choosing some special eigenvalues and eigenfunctions according to
the reduction conditions $q^{[n]}=-(r^{[n]})^*$, the determinant
representation  of $q^{[n]}$ provides some new solutions of the GI
equation. As examples, the breather solutions and rogue wave of the
GI is given explicitly by two-fold DT from a periodic ``seed" with a
constant amplitude.
\end{abstract}

\mbox{\hspace{3cm}}   breather solution, rogue wave.

{\bf PACS(2010) numbers}: 02.30.Ik, 42.81.Dp, 52.35.Bj, 52.35.Sb, 94.05.Fg




\section{Introduction}

Ablowitz-Kaup-Newell-Segur(AKNS) system\cite{AKNS} is one of the most important integrable systems
associated with a linear spectrum problem, which consists of well known nonlinear Schr\"odinger (NLS)
equation\cite{zs}. So it is a natural extension to consider a polynomial spectral problem(PSP) of
arbitrary order\cite{Konopelchenko}. Note that the corresponding Kirillov-Kostant symplectic form of
the general PSP is degenerated, which leads to some special restrictions  of the PSP in order to get
interesting integrable partial differential equations. The quadratic PSP under different
restrictions implies three kinds of the derivative nonlinear Schr\"odinger equations, which are
called DNLSI equation, DNLSII equation and DNLSIII equation respectively. The DNLSI equation \cite{KN} is given by
\begin{equation}\label{dnlsI}
iq_{t}-q_{xx}+i(q^2q^\ast)_{x}=0,
\end{equation}
and the DNLSII equation \cite{CL} is given by
\begin{equation}\label{CL}
iq_{t}+q_{xx}+iqq^\ast q_{x}=0,
\end{equation}
and the DNLSIII equation is in the form of\cite{GI}
\begin{equation}\label{GI}
iq_{t}+q_{xx}-iq^2q^{\ast}_{x}+\frac{1}{2}q^3{q^\ast}^2=0.
\end{equation}
The last equation is firstly found by Gerdjikov and Ivanov in
reference\cite{GI}, then it is also called GI equation. Here
asterisk denotes the complex conjugation, and subscript of $x$ (or
$t$) denotes the partial derivative with respect to $x$ (or $t$). On
the other hand, GI equation also can be regarded as an extension of
the NLS when certain higher-order nonlinear effects are taken into
account.

    Comparing with the intensive studies on the DNLSI \cite{KN,MDNLS,WDNLS,anderson,kenji1,steduel,
ruderman1,DNLStruncation,landaudamping,kawata1,cheng1,lashkin1,ruderman2,zhou,xuhe}
from the point of view of physics and mathematics, there are only
few works on the GI including soliton constructed by Darboux
transformation(DT)\cite{fan1}, Hamiltonian structures\cite{fan2},
algebro-geometric solutions\cite{fandai}, hierarchy of the GI
equation from an extended version of Drinfel'd-Sokolov
formulation\cite{Kikuchia}, Wronskian type solution\cite{Kikuchib}
without using affine Lie groups. In order to show more possible
physical relevance of the GI equation, and inspired by the
importance of breather(BA) solution and rogue wave(RW) of the
NLS\cite{Peregrine,Solli,Akhmediev1,Akhmediev2,Akhmediev3,Akhmediev4,matveevb,Akhmediev5},
so we shall find these two types  of solutions for the GI equation
by DT. Moreover, We also want to show the more differences between
DNLSI\cite{KN} and GI\cite{GI} through solutions and DT.

The organization of this paper is as follows. In section 2, it provides a relatively simple
approach to construct DT for the GI system, and then the determinant representation of the n-fold
DT  and  formulae of $q^{[n]}$ and $r^{[n]}$ are given  by eigenfunctions of spectral problem. The reduction of DT
of the GI system to the GI equation is also discussed by choosing
paired eigenvalues and eigenfunctions. In section 3, under specific reduction conditions, two
types of particular solutions-BA and RW are given by DT from a periodic ``seed"
 with a constant amplitude. The conclusions and discussions will be given in section 4.
\section{Darboux transformation}

Let us start from the first non-trivial flow of  a special quadratic spectral
problem\cite{GI},
\begin{equation}\label{sy1}
{\rm i}q_t+q_{xx}-{\rm i}q^2r_x+\frac{1}{2}q^3r^2=0,\\
\end{equation}
\begin{equation}\label{sy2}
{\rm i}r_t-r_{xx}+{\rm i}r^2q_x-\frac{1}{2}q^2r^3=0.\\
\end{equation}
which are  exactly reduced to the GI eq.(\ref{dnlsI}) for $r=-q^\ast$ while the choice $r
=q^\ast$ would lead to eq.(\ref{GI}) with the sign of the nonlinear term changed. The Lax pairs
corresponding to coupled GI equations(\ref{sy1}) and (\ref{sy2}) can be given by the
GI system with a quadratic spectral \cite{GI}
\begin{equation}\label{sys11}
 \pa_{x}\psi=(J\lambda^2+Q_{1}\lambda+Q_{0})\psi=U\psi,
\end{equation}
\begin{equation}\label{sys22}
\pa_{t}\psi=(2J\lambda^4+V_{3}\lambda^3+V_{2}\lambda^2+V_{1}\lambda+V_{0})\psi=V\psi.
\end{equation}
Here
\begin{equation}\label{fj1}
    \psi=\left( \begin{array}{c}
      \phi \\
      \varphi\\
     \end{array} \right),\nonumber\\
   \quad J= \left( \begin{array}{cc}
      -i &0 \\
      0 &i\\
   \end{array} \right),\nonumber\\
  \quad Q_{1}=\left( \begin{array}{cc}
     0 &q \\
     r &0\\
  \end{array} \right),\nonumber\\
  \quad Q_{0}=\left( \begin{array}{cc}
     -\dfrac{1}{2}i q r &0 \\
     0 &\dfrac{1}{2}i q r\\
  \end{array} \right),\nonumber\\
\end{equation}
\begin{equation}\label{fj2}
V_{3}=2Q_{1},  \quad V_{2}=Jqr, \quad V_{1}=\left( \begin{array}{cc}
0 &iq_{x} \\
 -ir_{x}&0\\
\end{array} \right),\quad V_{0}=\left( \begin{array}{cc}
\dfrac{1}{4}iq^2 r^2 &0 \\
 0&-\dfrac{1}{4}iq^2 r^2\\
\end{array} \right),\nonumber\\
\end{equation}
the spectral parameter $\lambda$ is an arbitrary complex constant and its eigenfunction is $\psi$.
Equations(\ref{sy1}) and (\ref{sy2}) are equivalent to the integrability condition  $U_{t}-V_{x}+[U,V]=0$ of
(\ref{sys11}) and (\ref{sys22}).

The main task of this section is to present a detailed derivation of the Darboux transforation of the
GI system and the determinant representation of the n-fold transformation.
Based on the DT for the NLS\cite{GN,matveev,he} and the DNLSI\cite{kenji1,steduel,xuhe},
 the main steps are : 1) to find a $2\times 2$ matrix  $T$ so that the GI spectral problem
 eq.(\ref{sys11}) and eq.(\ref{sys22}) is covariant, then get new
 solution $(q^{[1]},r^{[1]})$ expressed by elements of $T$ and
 seed solution $(q,r)$; 2) to find expressions of elements of $T$ in terms of eigenfunctions of
 GI spectral problem corresponding to the seed solution $(q,r)$; 3) to get the determinant
 representation of n-fold DT $T_n $ and new solutions $(q^{[n]},r^{[n]})$ by  $n$-times iteration of
  the DT; 4) to consider the reduction condition: $q^{[n]}= -(r^{[n]})^*$ by choosing special
  eigenvalue $\lambda_k$ and its eigenfunction $\psi_k$, and then get $q^{[n]}$ of the GI equation
  expressed by its seed solution $q$ and its associated eigenfunctions $\{\psi_k,k=1,2,\cdots,n\}$.
  However, we shall use the kernel of n-fold DT($T_n$) to fix it in the third step instead of
  iteration.

It is easy to see that the spectral problem (\ref{sys11}) and (\ref{sys22}) are transformed
to
  \begin{equation}\label{bh1}
{\psi^{[1]}}_{x}=U^{[1]}~\psi^{[1]},\ \ U^{[1]}=(T_{x}+T~U)T^{-1}.
\end{equation}
\begin{equation}\label{bh2}
{\psi^{[1]}}_{t}=V^{[1]}~\psi^{[1]}, \ \ V^{[1]}=(T_{t}+T~V)T^{-1}.
\end{equation}\\
under a gauge transformation \begin{equation}\label{bh3}
\psi^{[1]}=T~\psi.
\end{equation}
By cross differentiating (\ref{bh1}) and (\ref{bh2}), we obtain
\begin{equation}\label{bh4}
{U^{[1]}}_{t}-{V^{[1]}}_{x}+[{U^{[1]}},{V^{[1]}}]=T(U_{t}-V_{x}+[U,V])T^{-1}.
\end{equation}
This implies that, in order to make eqs.(\ref{sy1}) and eq.(\ref{sy2}) invariant under the
transformation (\ref{bh3}), it is crucial to search a matrix $T$ so that  $U^{[1]}$, $V^{[1]}$have
the same forms as $U$, $V$. At the same time the old potential(or
seed solution)($q$, $r$) in spectral matrixes $U$, $V$ are mapped
into new potentials
(or new solution)($q^{[1]}$, $r^{[1]}$) in transformed spectral matrixes $U^{[1]}$, $V^{[1]}$.\\

2.1 One-fold Darboux transformation of the GI system

Considering the universality of DT, suppose that the trial Darboux matrix $T$ in eq.(\ref{bh3}) is of
form
\begin{equation}\label{tt1}
T=T(\lambda)=\left( \begin{array}{cc}
a_{1}&b_{1} \\
c_{1} &d_{1}\\
\end{array} \right)\lambda+\left( \begin{array}{cc}
a_{0}&b_{0}\\
c_{0} &d_{0}\\
\end{array} \right),
\end{equation}
where $a_{0},  b_{0},  c_{0},   d_{0},   a_{1},   b_{1},   c_{1},   d_{1}$ are functions of $x$, $t$
to need be determined.
From \begin{equation}\label{tt2} T_{x}+T~U=U^{[1]}~T,
\end{equation}
comparing the coefficients of $\lambda^{j}, j=3, 2, 1, 0$, it yields
\begin{eqnarray}\label{xx1}
&\mbox{\hspace{-1cm}}&\lambda^{3}: b_{1}=0,\ c_{1}=0,\nonumber\\
&&\lambda^{2}: q~a_{1}+2~i~b_{0}-q^{[1]}d_{1}=0,\ -r^{[1]}~a_{1}+r~d_{1}-2~i~c_{0}=0,\nonumber\\
&&\lambda^{1}: {a_{1}}_{x}-\frac{1}{2}ia_{1}q r +rb_{0}+\dfrac{1}{2}iq^{[1]}r^{[1]}a_{1}-q^{[1]}c_{0}=0,\ {d_{1}}_{x}+\frac{1}{2}id_{1}q r -r^{[1]}b_{0}-\dfrac{1}{2}iq^{[1]}r^{[1]}d_{1}+qc_{0}=0,\nonumber\\
&&~~~~~~~q a_{0}-q^{[1]}d_{0}=0,-r^{[1]}a_{0}+r d_{0}=0,\nonumber\\
&&\lambda^{0}: {a_{0}}_{x}-\dfrac{1}{2}ia_{0}q r+\frac{1}{2}iq^{[1]}r^{[1]}a_{0}=0,{b_{0}}_{x}+\dfrac{1}{2}ib_{0}q r+\frac{1}{2}iq^{[1]}r^{[1]}b_{0}=0,\nonumber\\
&\mbox{\hspace{-1cm}}&~~~~~~~{c_{0}}_{x}-\dfrac{1}{2}ic_{0}q r-\frac{1}{2}iq^{[1]}r^{[1]}c_{0}=0,
{d_{0}}_{x}+\dfrac{1}{2}id_{0}q r-\frac{1}{2}iq^{[1]}r^{[1]}d_{0}=0.
\end{eqnarray}

Similarly, from  \begin{equation}\label{tt3} T_{t}+T~V=V^{[1]}~T,
\end{equation} comparing the coefficients of $\lambda^{j} ,j =4,3, 2, 1, 0$,it implies
\begin{eqnarray}\label{ttt1}
&&\lambda^{4}: 2 i b_{0}-q^{[1]}d_{1}+q a_{1}=0,\ -2 i c_{0}-2r^{[1]}a_{1}+r d_{1}=0,\nonumber\\
&&\lambda^{3}: r^{[1]} q^{[1]} a_{1} i-2 q^{[1]} c_{0}-a_{1} r q i+2 r b_{0}=0,\ q a_{0}-q^{[1]}d_{0}=0,\nonumber\\
&&~~~~~~~~r d_{0}-r^{[1]}a_{0}=0,d_{1} r q i-r^{[1]}q^{[1]}d_{1}i+2qc_{0}-2r^{[1]}b_{0}=0,\nonumber\\
&&\lambda^{2}: a_{0} r q-a_{0} r^{[1]}q^{[1]}=0,\ -b_{0} r q i +q^{[1]}_{x} d_{1} i-a_{1} q_{x} i- r^{[1]} q^{[1]} b_{0} i=0,\nonumber\\
&&~~~~~~~~c_{0} r q i+r^{[1]} q^{[1]} c_{0} i+d_{1} r_{x} i-{r^{[1]}}_{x}a_{1} i=0,\ r^{[1]}q^{[1]}d_{0}-r q d_{0}=0,\nonumber\\
&&\lambda^{1}: {a_{1}}_{t}-{q^{[1]}}_{x} c_{0} i-b_{0} r_{x}i+\dfrac{1}{4} a_{1}r^{2} q^{2}-\dfrac{1}{4} a_{1} {r^{[1]}}^{2}{q^{[1]}}^{2}-\dfrac{1} {2}a_{1}q r_{x} +\dfrac{1}{2}a_{1}r q_{x} +\dfrac{1}{2}a_{1}q^{[1]}{r^{[1]}}_{x}-\dfrac{1} {2}a_{1}r^{[1]} {q^{[1]}}_{x}=0,\nonumber\\
&&~~~~~~~~ {d_{1}}_{t}+q_{x} c_{0} i+b_{0} {r^{[1]}}_{x}i     -\dfrac{1}{4} d_{1}r^{2} q^{2}+\dfrac{1}{4} d_{1} {r^{[1]}}^{2}{q^{[1]}}^{2}  +\dfrac{1} {2}d_{1}  q r_{x} -\dfrac{1} {2}d_{1}  r q_{x} -\dfrac{1} {2}d_{1}  q^{[1]} {r^{[1]}}_{x} +\dfrac{1} {2}d_{1}   r^{[1]} {q^{[1]}}_{x}  =0,\nonumber\\
&&~~~~~~~~ {q^{[1]}}_{x} d_{0} i-a_{0} q_{x}i=0,\ d_{0} r_{x}i-{r^{[1]}}_{x} a_{0} i=0, \nonumber\\
&&\lambda^{0}: {a_{0}}_{t}+\dfrac{1}{4} a_{0}r^{2} q^{2}-\dfrac{1}{4} a_{0} {r^{[1]}}^{2}{q^{[1]}}^{2}  -\dfrac{1} {2}a_{0}  q r_{x} +\dfrac{1} {2}a_{0}  r q_{x} +\dfrac{1} {2}a_{0}  q^{[1]} {r^{[1]}}_{x} -\dfrac{1} {2}a_{0}   r^{[1]} {q^{[1]}}_{x}=0,\nonumber\\
&&~~~~~~~~{b_{0}}_{t}-\dfrac{1}{4} b_{0}r^{2} q^{2}-\dfrac{1}{4} b_{0} {r^{[1]}}^{2}{q^{[1]}}^{2}  +\dfrac{1} {2}b_{0}  q r_{x} -\dfrac{1} {2}b_{0}  r q_{x} -\dfrac{1} {2}b_{0}  q^{[1]} {r^{[1]}}_{x} +\dfrac{1} {2}b_{0}   r^{[1]} {q^{[1]}}_{x}=0,\nonumber\\
&&~~~~~~~~{c_{0}}_{t}+\dfrac{1}{4} c_{0}r^{2} q^{2}+\dfrac{1}{4} c_{0} {r^{[1]}}^{2}{q^{[1]}}^{2}  -\dfrac{1} {2}c_{0}  q r_{x} +\dfrac{1} {2}c_{0}  r q_{x} +\dfrac{1} {2}c_{0}  q^{[1]} {r^{[1]}}_{x} -\dfrac{1} {2}c_{0}   r^{[1]} {q^{[1]}}_{x}=0,\nonumber\\
&&~~~~~~~~{d_{0}}_{t}-\dfrac{1}{4} d_{0}r^{2} q^{2}-\dfrac{1}{4} d_{0} {r^{[1]}}^{2}{q^{[1]}}^{2}  +\dfrac{1} {2}d_{0}  q r_{x} -\dfrac{1} {2}d_{0}  r q_{x} -\dfrac{1} {2}d_{0}  q^{[1]} {r^{[1]}}_{x} +\dfrac{1} {2}d_{0}   r^{[1]} {q^{[1]}}_{x}=0.
\end{eqnarray}
Note  that the coefficient of $\lambda^5$ in eq.(\ref{tt3}) is an identity automatically
because of the first formula of eq.(\ref{xx1})

In order to get the non-trivial solutions, we shall construct a
basic(or one-fold ) Darboux transformation matrix $T$ under an
assumption $ a_{0} = 0, d_{0} = 0$. If we set $ a_{0} \not=0$, then
$d_{0}$ is not zero,
 and furthermore find that some coefficients $(a_{1},d_{1},a_{0},d_{0})$ of $T$ are constants
 except  $b_{0}=\dfrac{1}{2}i q (\dfrac{-a_{0}d_{1}}{d_{0}}+a_{1})$ and
 $c_{0}=\dfrac{1}{2}i r (\dfrac{a_{1}d_{0}}{a_{0}}-d_{1})$, which gives a trivial DT.
 This means our assumption does not decrease the generality of the DT of the GI system. Based on
 eq.(\ref{xx1}), eq.(\ref{ttt1}) and above analysis,
let  Darboux matrix $T$ be  the  form  of
\begin{equation}\label{TT}
 T_{1}=T_{1}(\lambda;\lambda_1)=\left( \begin{array}{cc}
a_{1}&0 \\
0 &d_{1}\\
\end{array} \right)\lambda+\left( \begin{array}{cc}
0&b_{0}\\
c_{0} &0\\
\end{array} \right),
\end{equation}
without losing any generality. Here $b_{0}, c_{0}$ are undetermined
function of ($x$, $t$), which will be expressed by the eigenfunction
associated with $\lambda_1$ in the GI spectral problem. However,
$a_1$ and $d_1$ are constants, which will be verified later. First
of all, we introduce $n$ eigenfunctions $\psi_j$ as
\begin{eqnarray}
&&\psi_{j}=\left(
\begin{array}{c}\label{jie2}
 \phi_{j}   \\
 \varphi_{j}  \\
\end{array} \right),\ \ j=1,2,....n,\phi_{j}=\phi_{j}(x,t,\lambda_{j}), \
\varphi_{j}=\varphi_{j}(x,t,\lambda_{j}). \label{jie1}
\end{eqnarray}

\noindent {\bf Theorem 1.}{\sl  The  elements of one-fold DT  are  parameterized by the
eigenfunction $\psi_1$ associated with
$\lambda_1$ as
 \begin{eqnarray}
a_{1}=d_{1}=1, \ \ b_{0}=-\frac{\lambda_{1}\phi_{1}}{\varphi_{1}},
\ \ c_{0}=-\frac{\lambda_{1}\varphi_{1}}{\phi_{1}}, \label{DT1aibi}  \\
\Leftrightarrow
T_1(\lambda;\lambda_1)=\left(
\begin{array}{cc}
\lambda & -\frac{\lambda_{1}\phi_{1}}{\varphi_{1}}\\
-\frac{\lambda_{1}\varphi_{1}}{\phi_{1}} & \lambda \\
\end{array}  \right),  \label{DT1matrix}
\end{eqnarray}
and then the new solutions $q^{[1]}$ and $r^{[1]}$ are given by
\begin{eqnarray}\label{sTT}
q^{[1]}=q-2i\frac{\lambda_{1}\phi_{1}}{\varphi_{1}},
r^{[1]}=r+2i\frac{\lambda_{1}\varphi_{1}}{\phi_{1}},
\end{eqnarray}
and the new eigenfunction $\psi_j^{[1]}$ corresponding to $\lambda_j$ is
\begin{equation}
\psi^{[1]}_j=
\left(
\begin{array}{c}
\dfrac{1}{\varphi_1}\left|\begin{array}{cc}
 \varphi_j&-\lambda_j\phi_j \\
\varphi_1&-\lambda_1\phi_1
 \end{array}\right| \\ \\
\dfrac{1}{\phi_1}
\left|\begin{array}{cc}
 \phi_j&-\lambda_j\varphi_j \\
 \phi_1&-\lambda_1\varphi_1
 \end{array}\right|
\end{array}
\right).
\end{equation}
}
{\bf Proof.} Note that $(b_{0}c_{0})_{x}=0$, ${a_{1}}_{x}=0$ and ${d_{1}}_{x}=0$ is
derived from the eq.(\ref{xx1}), then  $(b_{0}c_{0})_{t}=0$ , ${a_{1}}_{t}=0$ and ${d_{1}}_{t}=0$
 is derived from the eq.(\ref{ttt1}). So
$a_{1}$ and $d_{1} $ are constants, and then let $a_{1}=d_{1}=1$ for simplicity of later
calculations. By transformation eq.(\ref{TT}) and eq.(\ref{xx1}), new solutions
are given by
 \begin{eqnarray} \label{TT1}
q^{[1]}=\dfrac{a_{1}}{d_{1}}q+2~i~\dfrac{b_{0}}{d_{1}},r^{[1]}=\dfrac{d_{1}}{a_{1}}q-2~i~\dfrac{c_{0}}{a_{1}}.
\end{eqnarray}
By using a general fact of the DT, i.e.,
 $T_{1}(\lambda;\lambda_{1})|_{\lambda=\lambda_1}\psi_{1}=0$, then eq.(\ref{DT1aibi}) is obtained.
Next, substituting $(a_1,d_1,b_0,c_0)$ given in eq.(\ref{DT1aibi})
into eq.(\ref{TT1}), then new solutions are given as eq.
(\ref{sTT}). Further, by using the explicit matrix representation
eq.(\ref{DT1matrix}) of $T_1$, then $\psi^{[1]}_j$ is given by
\begin{equation}
\psi^{[1]}_j=T_1(\lambda;\lambda_1)|_{\lambda=\lambda_j} \psi_j=\left.\left(
\begin{array}{cc}
\lambda & -\frac{\lambda_{1}\phi_{1}}{\varphi_{1}}\\
-\frac{\lambda_{1}\varphi_{1}}{\phi_{1}} & \lambda \\
\end{array}  \right)\right|_{\lambda=\lambda_j} \left( \begin{array} {c}
\phi_j\\
\varphi_j
\end{array} \right)=
\left(
\begin{array}{c}
\dfrac{1}{\varphi_1}\left|\begin{array}{cc}
 \varphi_j&-\lambda_j\phi_j \\
\varphi_1&-\lambda_1\phi_1
 \end{array}\right| \\ \\
\dfrac{1}{\phi_1}
\left|\begin{array}{cc}
 \phi_j&-\lambda_j\varphi_j \\
 \phi_1&-\lambda_1\varphi_1
 \end{array}\right|
\end{array}
\right).
\end{equation}
Last, a tedious calculation shows that $T_1$ in eq.(\ref{DT1matrix}) and new solutions
indeed satisfy eq.(\ref{tt3}) or (equivalently eq.(\ref{ttt1})). So
GI spectral  problem  is  covariant under transformation $T_1$ in eq.(\ref{DT1matrix}) and
eq.(\ref{sTT}), and thus it is the DT of eq.(\ref{sy1}) and eq.(\ref{sy2}). $\square$

2.2 N-fold Darboux transformation for GI system

The key task is to establish the determinant representation of the n-fold DT
for GI system in this subsection. To this purpose, set
$$
\begin{array}{cc}
\textbf{D}=&\left\{\left.\left(\begin{array}{cc}
a& 0\\
0&d
  \end{array} \right)\right| a,d \text{ are complex functions of}\ x\ \text{and}\ t  \right\},\\
\textbf{A}=&\left\{\left.\left(\begin{array}{cc}
0& b\\
c&0
  \end{array} \right)\right| b,c \text{ are complex functions of}\ x\ \text{and}\ t  \right\},
\end{array}
$$
as ref.\cite{kenji1}.

 According to the form of $T_1$ in eq.(\ref{TT}), the n-fold DT should be the form of
\begin{equation}\label{tnss}T_{n}=T_{n}(\lambda;\lambda_1,\lambda_2, \cdots,\lambda_n)
=\sum_{l=0}^{n}P_{l}\lambda^{l},
\end{equation}
with \begin{eqnarray}
\label{tnsss}
P_{n}=\left( \mbox{\hspace{-0.2cm}}
\begin{array}{cc}
1\mbox{\hspace{-0.1cm}}&0 \\
0 \mbox{\hspace{-0.1cm}}&1\\
\end{array}  \mbox{\hspace{-0.2cm}}\right)\in \textbf{D},\ P_{n-1}=\left( \mbox{\hspace{-0.2cm}}
\begin{array}{cc}
0 \mbox{\hspace{-0.3cm}}&b_{n-1} \\
c_{n-1} \mbox{\hspace{-0.3cm}}&0\\
\end{array}  \mbox{\hspace{-0.2cm}}\right)\in \textbf{A},\ P_{l}\in \textbf{D}&\textrm{\small \mbox{\hspace{-0.3cm}}(if $l-n$ is even)}
,\ P_{l}\in \textbf{A}&\textrm{\small \mbox{\hspace{-0.3cm}}(if $l-n$ is odd)}.
\end{eqnarray}
Here  $P_i(0\leq i\leq n-1)$ is the function of $x$ and $t$.
In particular, $P_0 \in \textbf{D}$ if $n$ is even and $P_0 \in \textbf{A}$ if $n$ is odd, which
leads to the separate discussion on the determinant representation of $T_n$ in the following
by means of its kernel. Specifically, from algebraic equations,
\begin{equation}\label{ttnss}
\psi_{k}^{[n]}=T_{n}(\lambda;\lambda_1,\cdots,\lambda_n)|_{\lambda=\lambda_k}\psi_{k}=\sum_{l=0}^{n}P_{l}\lambda_{k}^{l}\psi_{k}=0,
k=1,2,\cdots,n,
\end{equation}
coefficients $P_i$ is solved by Cramer's rule. Thus we get determinant representation of the $T_n$.

\noindent {\bf Theorem2.} (1)For $n=2k(k=1,2,3,\cdots)$, the n-fold DT of the GI system can be expressed by
\begin{equation}
\label{fss1}T_{n}=T_{n}(\lambda;\lambda_1,\lambda_2,\cdots,\lambda_n)=\left(
\begin{array}{cc}
\dfrac{\widetilde{(T_{n})_{11}}}{W_{n}}& \dfrac{\widetilde{(T_{n})_{12}}}{W_{n}}\\ \\
\dfrac{\widetilde{(T_{n})_{21}}}{\widetilde{W_{n}}}& \dfrac{\widetilde{(T_{n})_{22}}}{\widetilde{W_{n}}}\\
\end{array} \right),
\end{equation}
with
\begin{equation}\label{fsst1}
W_{n}=\begin{vmatrix}
\phi_{1}&\lambda_{1}\varphi_{1}&\ldots&\lambda_{1}^{n-3}\varphi_{1}&\lambda_{1}^{n-2}\phi_{1}&\lambda_{1}^{n-1}\varphi_{1}\\
\phi_{2}&\lambda_{2}\varphi_{2}&\ldots&\lambda_{2}^{n-3}\varphi_{2}&\lambda_{2}^{n-2}\phi_{2}&\lambda_{2}^{n-1}\varphi_{2}\\
\vdots&\vdots&\vdots&\vdots&\vdots&\vdots\\
\phi_{n}&\lambda_{n}\varphi_{n}&\ldots&\lambda_{n}^{n-3}\varphi_{n}&\lambda_{n}^{n-2}\phi_{n}&\lambda_{n}^{n-1}\varphi_{n}\nonumber\\
\end{vmatrix},
\end{equation}
\begin{equation}\label{fsst2}
\widetilde{(T_{n})_{11}}=\begin{vmatrix}
1&0&\ldots&0&\lambda^{n-2}&0&\lambda^{n}\\
\phi_{1}&\lambda_{1}\varphi_{1}&\ldots&\lambda_{1}^{n-3}\varphi_{1}&\lambda_{1}^{n-2}\phi_{1}&\lambda_{1}^{n-1}\varphi_{1}&\lambda_{1}^{n}\phi_{1}\\
\phi_{2}&\lambda_{2}\varphi_{2}&\ldots&\lambda_{2}^{n-3}\varphi_{2}&\lambda_{2}^{n-2}\phi_{2}&\lambda_{2}^{n-1}\varphi_{2}&\lambda_{2}^{n}\phi_{2}\\
\vdots&\vdots&\vdots&\vdots&\vdots&\vdots&\vdots\\
\phi_{n}&\lambda_{n}\varphi_{n}&\ldots&\lambda_{n}^{n-3}\varphi_{n}&\lambda_{n}^{n-2}\phi_{n}&\lambda_{n}^{n-1}\varphi_{n}&\lambda_{n}^{n}\phi_{n}\nonumber\\
\end{vmatrix},
\end{equation}
\begin{equation}\label{fsst3}
\widetilde{(T_{n})_{12}}=\begin{vmatrix}
0&\lambda&\ldots&\lambda^{n-3}&0&\lambda^{n-1}&0\\
\phi_{1}&\lambda_{1}\varphi_{1}&\ldots&\lambda_{1}^{n-3}\varphi_{1}&\lambda_{1}^{n-2}\phi_{1}&\lambda_{1}^{n-1}\varphi_{1}&\lambda_{1}^{n}\phi_{1}\\
\phi_{2}&\lambda_{2}\varphi_{2}&\ldots&\lambda_{2}^{n-3}\varphi_{2}&\lambda_{2}^{n-2}\phi_{2}&\lambda_{2}^{n-1}\varphi_{2}&\lambda_{2}^{n}\phi_{2}\\
\vdots&\vdots&\vdots&\vdots&\vdots&\vdots&\vdots\\
\phi_{n}&\lambda_{n}\varphi_{n}&\ldots&\lambda_{n}^{n-3}\varphi_{n}&\lambda_{n}^{n-2}\phi_{n}&\lambda_{n}^{n-1}\varphi_{n}\nonumber&\lambda_{n}^{n}\phi_{n}\nonumber\\
\end{vmatrix},
\end{equation}
\begin{equation}\label{fsst4}
\widetilde{W_{n}}=\begin{vmatrix}
\varphi_{1}&\lambda_{1}\phi_{1}&\ldots&\lambda_{1}^{n-3}\phi_{1}&\lambda_{1}^{n-2}\varphi_{1}&\lambda_{1}^{n-1}\phi_{1}\\
\varphi_{2}&\lambda_{2}\phi_{2}&\ldots&\lambda_{2}^{n-3}\phi_{2}&\lambda_{2}^{n-2}\varphi_{2}&\lambda_{2}^{n-1}\phi_{2}\\
\vdots&\vdots&\vdots&\vdots&\vdots&\vdots\\
\varphi_{n}&\lambda_{n}\phi_{n}&\ldots&\lambda_{n}^{n-3}\phi_{n}&\lambda_{n}^{n-2}\varphi_{n}&\lambda_{n}^{n-1}\phi_{n}\nonumber\\
\end{vmatrix},
\end{equation}
\begin{equation}\label{fsst5}
\widetilde{(T_{n})_{21}}=\begin{vmatrix}
0&\lambda&\ldots&\lambda^{n-3}&0&\lambda^{n-1}&0\\
\varphi_{1}&\lambda_{1}\phi_{1}&\ldots&\lambda_{1}^{n-3}\phi_{1}&\lambda_{1}^{n-2}\varphi_{1}&\lambda_{1}^{n-1}\phi_{1}&\lambda_{1}^{n}\varphi_{1}\\
\varphi_{2}&\lambda_{2}\phi_{2}&\ldots&\lambda_{2}^{n-3}\phi_{2}&\lambda_{2}^{n-2}\varphi_{2}&\lambda_{2}^{n-1}\phi_{2}&\lambda_{2}^{n}\varphi_{2}\\
\vdots&\vdots&\vdots&\vdots&\vdots&\vdots&\vdots\\
\varphi_{n}&\lambda_{n}\phi_{n}&\ldots&\lambda_{n}^{n-3}\phi_{n}&\lambda_{n}^{n-2}\varphi_{n}&\lambda_{n}^{n-1}\phi_{n}&\lambda_{n}^{n}\varphi_{n}\nonumber\\
\end{vmatrix},
\end{equation}
\begin{equation}\label{fsst6}
\widetilde{(T_{n})_{22}}=\begin{vmatrix}
1&0&\ldots&0&\lambda^{n-2}&0&\lambda^{n}\\
\varphi_{1}&\lambda_{1}\phi_{1}&\ldots&\lambda_{1}^{n-3}\phi_{1}&\lambda_{1}^{n-2}\varphi_{1}&\lambda_{1}^{n-1}\phi_{1}&\lambda_{1}^{n}\varphi_{1}\\
\varphi_{2}&\lambda_{2}\phi_{2}&\ldots&\lambda_{2}^{n-3}\phi_{2}&\lambda_{2}^{n-2}\varphi_{2}&\lambda_{2}^{n-1}\phi_{2}&\lambda_{2}^{n}\varphi_{2}\\
\vdots&\vdots&\vdots&\vdots&\vdots&\vdots&\vdots\\
\varphi_{n}&\lambda_{n}\phi_{n}&\ldots&\lambda_{n}^{n-3}\phi_{n}&\lambda_{n}^{n-2}\varphi_{n}&\lambda_{n}^{n-1}\phi_{n}&\lambda_{n}^{n}\varphi_{n}\nonumber\\
\end{vmatrix}.
\end{equation}
(2)For $n=2k+1(k=1,2,3,\cdots)$, then
\begin{equation}\label{fss2}
T_{n}=T_{n}(\lambda;\lambda_1,\lambda_2,\cdots,\lambda_n)=\left(
\begin{array}{cc}
\dfrac{\widehat{(T_{n})_{11}}}{Q_{n}}& \dfrac{\widehat{(T_{n})_{12}}}{Q_{n}}\\ \\
\dfrac{\widehat{(T_{n})_{21}}}{\widehat{Q_{n}}}& \dfrac{\widehat{(T_{n})_{22}}}{\widehat{Q_{n}}}\\
\end{array} \right),
\end{equation}
with
\begin{equation}\label{fsstt1}
Q_{n}=\begin{vmatrix}
\varphi_{1}&\lambda_{1}\phi_{1}&\ldots&\lambda_{1}^{n-3}\phi_{1}&\lambda_{1}^{n-2}\varphi_{1}&\lambda_{1}^{n-1}\varphi_{1}\\
\varphi_{2}&\lambda_{2}\phi_{2}&\ldots&\lambda_{2}^{n-3}\phi_{2}&\lambda_{2}^{n-2}\varphi_{2}&\lambda_{2}^{n-1}\varphi_{2}\\
\vdots&\vdots&\vdots&\vdots&\vdots&\vdots\\
\varphi_{n}&\lambda_{n}\phi_{n}&\ldots&\lambda_{n}^{n-3}\phi_{n}&\lambda_{n}^{n-2}\varphi_{n}&\lambda_{n}^{n-1}\varphi_{n}\nonumber\\
\end{vmatrix},
\end{equation}
\begin{equation}\label{fsstt2}
\widehat{(T_{n})_{11}}=\begin{vmatrix}
0&\lambda&\ldots&0&\lambda^{n-2}&0&\lambda^{n}\\
\varphi_{1}&\lambda_{1}\phi_{1}&\ldots&\lambda_{1}^{n-3}\phi_{1}&\lambda_{1}^{n-2}\varphi_{1}&\lambda_{1}^{n-1}\varphi_{1}&-\lambda_{1}^{n}\phi_{1}\\
\varphi_{2}&\lambda_{2}\phi_{2}&\ldots&\lambda_{2}^{n-3}\phi_{2}&\lambda_{2}^{n-2}\varphi_{2}&\lambda_{2}^{n-1}\varphi_{2}&-\lambda_{2}^{n}\phi_{2}\\
\vdots&\vdots&\vdots&\vdots&\vdots&\vdots&\vdots\\
\varphi_{n}&\lambda_{n}\phi_{n}&\ldots&\lambda_{n}^{n-3}\phi_{n}&\lambda_{n}^{n-2}\varphi_{n}&\lambda_{n}^{n-1}\varphi_{n}&\lambda_{n}^{n}\phi_{n}\nonumber\\
\end{vmatrix},
\end{equation}
\begin{equation}\label{fsstt3}
\widehat{(T_{n})_{12}}=\begin{vmatrix}
1&0&\ldots&\lambda^{n-3}&0&\lambda^{n-1}&0\\
\varphi_{1}&\lambda_{1}\phi_{1}&\ldots&\lambda_{1}^{n-3}\phi_{1}&\lambda_{1}^{n-2}\varphi_{1}&\lambda_{1}^{n-1}\varphi_{1}&-\lambda_{1}^{n}\phi_{1}\\
\varphi_{2}&\lambda_{2}\phi_{2}&\ldots&\lambda_{2}^{n-3}\phi_{2}&\lambda_{2}^{n-2}\varphi_{2}&\lambda_{2}^{n-1}\varphi_{2}&-\lambda_{2}^{n}\phi_{2}\\
\vdots&\vdots&\vdots&\vdots&\vdots&\vdots&\vdots\\
\varphi_{n}&\lambda_{n}\phi_{n}&\ldots&\lambda_{n}^{n-3}\phi_{n}&\lambda_{n}^{n-2}\varphi_{n}&\lambda_{n}^{n-1}\varphi_{n}&-\lambda_{n}^{n}\phi_{n}\nonumber\\
\end{vmatrix},
\end{equation}
\begin{equation}\label{fsstt4}
\widehat{Q_{n}}=\begin{vmatrix}
\phi_{1}&\lambda_{1}\varphi_{1}&\ldots&\lambda_{1}^{n-3}\varphi_{1}&\lambda_{1}^{n-2}\phi_{1}&\lambda_{1}^{n-1}\phi_{1}\\
\phi_{2}&\lambda_{2}\varphi_{2}&\ldots&\lambda_{2}^{n-3}\varphi_{2}&\lambda_{2}^{n-2}\phi_{2}&\lambda_{2}^{n-1}\phi_{2}\\
\vdots&\vdots&\vdots&\vdots&\vdots&\vdots\\
\phi_{n}&\lambda_{n}\varphi_{n}&\ldots&\lambda_{n}^{n-3}\varphi_{n}&\lambda_{n}^{n-2}\phi_{n}&\lambda_{n}^{n-1}\phi_{n}\nonumber\\
\end{vmatrix},
\end{equation}
\begin{equation}\label{fsstt5}
\widehat{(T_{n})_{21}}=\begin{vmatrix}
1&0&\ldots&\lambda^{n-3}&0&\lambda^{n-1}&0\\
\phi_{1}&\lambda_{1}\varphi_{1}&\ldots&\lambda_{1}^{n-3}\varphi_{1}&\lambda_{1}^{n-2}\phi_{1}&\lambda_{1}^{n-1}\phi_{1}&-\lambda_{1}^{n}\varphi_{1}\\
\phi_{2}&\lambda_{2}\varphi_{2}&\ldots&\lambda_{2}^{n-3}\varphi_{2}&\lambda_{2}^{n-2}\phi_{2}&\lambda_{2}^{n-1}\phi_{2}&-\lambda_{2}^{n}\varphi_{2}\\
\vdots&\vdots&\vdots&\vdots&\vdots&\vdots&\vdots\\
\phi_{n}&\lambda_{n}\varphi_{n}&\ldots&\lambda_{n}^{n-3}\varphi_{n}&\lambda_{n}^{n-2}\phi_{n}&\lambda_{n}^{n-1}\phi_{n}&\-\lambda_{n}^{n}\varphi_{n}\nonumber\\
\end{vmatrix},
\end{equation}
\begin{equation}\label{fsstt6}
\widehat{(T_{n})_{22}}=\begin{vmatrix}
0&\lambda&\ldots&0&\lambda^{n-2}&0&\lambda^{n}\\
\phi_{1}&\lambda_{1}\varphi_{1}&\ldots&\lambda_{1}^{n-3}\varphi_{1}&\lambda_{1}^{n-2}\phi_{1}&\lambda_{1}^{n-1}\phi_{1}&-\lambda_{1}^{n}\varphi_{1}\\
\phi_{2}&\lambda_{2}\varphi_{2}&\ldots&\lambda_{2}^{n-3}\varphi_{2}&\lambda_{2}^{n-2}\phi_{2}&\lambda_{2}^{n-1}\phi_{2}&-\lambda_{2}^{n}\varphi_{2}\\
\vdots&\vdots&\vdots&\vdots&\vdots&\vdots&\vdots\\
\phi_{n}&\lambda_{n}\varphi_{n}&\ldots&\lambda_{n}^{n-3}\varphi_{n}&\lambda_{n}^{n-2}\phi_{n}&\lambda_{n}^{n-1}\phi_{n}&\-\lambda_{n}^{n}\varphi_{n}\nonumber\\
\end{vmatrix}.
\end{equation}

Next, we consider the transformed new solutions ($q^{[n]},r^{[n]}$)of GI system corresponding to
the n-fold DT. Under covariant requirement of spectral problem of the GI system, the transformed
form should be
\begin{equation}\label{nsys11}
\pa_{x}\psi^{[n]}=(J\lambda^2+{Q_{1}}^{[n]}\lambda+{Q_{0}}^{[n]})\psi=U^{[n]}\psi,
\end{equation}
with
\begin{equation}\label{nfj1}
     \psi=\left( \begin{array}{c}
          \phi   \\
          \varphi \\
      \end{array} \right),
    \quad J= \left( \begin{array}{cc}
       i &0 \\
       0 &-i\\
     \end{array} \right),
  \quad {Q_{1}}^{[n]}=\left( \begin{array}{cc}
    0 &q^{[n]} \\
    r^{[n]} &0\\
\end{array} \right),
\quad {Q_{0}}^{[n]}=\left( \begin{array}{cc}
     -\dfrac{1}{2}i q^{[n]} r^{[n]} &0 \\
     0 &\dfrac{1}{2}i q^{[n]} r^{[n]}\\
  \end{array} \right),\nonumber\\
\end{equation}
and then
\begin{equation}\label{ntt2} {T_{n}}_{x}+T_{n}~U=U^{[n]}~T_{n}.
\end{equation}
Substituting $T_n$ given by eq.(\ref{tnss}) into eq.(\ref{ntt2}),and then comparing the
coefficients of $\lambda^{n+1}$, it yields
\begin{eqnarray}\label{ntt3}
&&q^{[n]}=q+2ib_{n-1},
\ \ r^{[n]}=r-2ic_{n-1}.
\end{eqnarray}
Furthermore, taking $b_{n-1},c_{n-1}$
which are obtained from eq.(\ref{fss1}) for $n=2k$ and
from eq.(\ref{fss2}) for $n=2k+1$, into (\ref{ntt3}), then new solutions ($q^{[n]},r^{[n]}$)
are given by
\begin{eqnarray}\label{ntt4}
&&q^{[n]}=q+2i\dfrac{\Omega_{11}}{\Omega_{12}}, \ \ r^{[n]}=r-2i\dfrac{\Omega_{21}}{\Omega_{22}}.
\end{eqnarray}
Here, (1)for $n=2k$,
\begin{equation}\label{ntt5}
\Omega_{11}=\begin{vmatrix}
\phi_{1}&\lambda_{1}\varphi_{1}&\ldots&\lambda_{1}^{n-3}\varphi_{1}&\lambda_{1}^{n-2}\phi_{1}&-\lambda_{1}^{n}\phi_{1}\\
\phi_{2}&\lambda_{2}\varphi_{2}&\ldots&\lambda_{2}^{n-3}\varphi_{2}&\lambda_{2}^{n-2}\phi_{2}&-\lambda_{2}^{n}\phi_{2}\\
\vdots&\vdots&\vdots&\vdots&\vdots&\vdots\\
\phi_{n}&\lambda_{n}\varphi_{n}&\ldots&\lambda_{n}^{n-3}\varphi_{n}&\lambda_{n}^{n-2}\phi_{n}&-\lambda_{n}^{n}\phi_{n}\\
\end{vmatrix},
\end{equation}
\begin{equation*}
\Omega_{12}=\begin{vmatrix}
\phi_{1}&\lambda_{1}\varphi_{1}&\ldots&\lambda_{1}^{n-3}\varphi_{1}&\lambda_{1}^{n-2}\phi_{1}&\lambda_{1}^{n-1}\varphi_{1}\\
\phi_{2}&\lambda_{2}\varphi_{2}&\ldots&\lambda_{2}^{n-3}\varphi_{2}&\lambda_{2}^{n-2}\phi_{2}&\lambda_{2}^{n-1}\varphi_{2}\\
\vdots&\vdots&\vdots&\vdots&\vdots&\vdots\\
\phi_{n}&\lambda_{n}\varphi_{n}&\ldots&\lambda_{n}^{n-3}\varphi_{n}&\lambda_{n}^{n-2}\phi_{n}&\lambda_{n}^{n-1}\varphi_{n}\\
\end{vmatrix},
\end{equation*}
\begin{equation*}
\Omega_{21}=\begin{vmatrix}
\varphi_{1}&\lambda_{1}\phi_{1}&\ldots&\lambda_{1}^{n-3}\phi_{1}&\lambda_{1}^{n-2}\varphi_{1}&-\lambda_{1}^{n}\varphi_{1}\\
\varphi_{2}&\lambda_{2}\phi_{2}&\ldots&\lambda_{2}^{n-3}\phi_{2}&\lambda_{2}^{n-2}\varphi_{2}&-\lambda_{2}^{n}\varphi_{2}\\
\vdots&\vdots&\vdots&\vdots&\vdots&\vdots\\
\varphi_{n}&\lambda_{n}\phi_{n}&\ldots&\lambda_{n}^{n-3}\phi_{n}&\lambda_{n}^{n-2}\varphi_{n}&-\lambda_{n}^{n}\varphi_{n}\\
\end{vmatrix},
\end{equation*}
\begin{equation*}
\Omega_{22}=\begin{vmatrix}
\varphi_{1}&\lambda_{1}\phi_{1}&\ldots&\lambda_{1}^{n-3}\phi_{1}&\lambda_{1}^{n-2}\varphi_{1}&\lambda_{1}^{n-1}\phi_{1}\\
\varphi_{2}&\lambda_{2}\phi_{2}&\ldots&\lambda_{2}^{n-3}\phi_{2}&\lambda_{2}^{n-2}\varphi_{2}&\lambda_{2}^{n-1}\phi_{2}\\
\vdots&\vdots&\vdots&\vdots&\vdots&\vdots\\
\varphi_{n}&\lambda_{n}\phi_{n}&\ldots&\lambda_{n}^{n-3}\phi_{n}&\lambda_{n}^{n-2}\varphi_{n}&\lambda_{n}^{n-1}\phi_{n}\\
\end{vmatrix};
\end{equation*}

(2) for $n=2k+1$,
\begin{equation}\label{ntt6}
\Omega_{11}=\begin{vmatrix}
\varphi_{1}&\lambda_{1}\phi_{1}&\ldots&\lambda_{1}^{n-3}\phi_{1}&\lambda_{1}^{n-2}\varphi_{1}&-\lambda_{1}^{n}\phi_{1}\\
\varphi_{2}&\lambda_{2}\phi_{2}&\ldots&\lambda_{2}^{n-3}\phi_{2}&\lambda_{2}^{n-2}\varphi_{2}&-\lambda_{2}^{n}\phi_{2}\\
\vdots&\vdots&\vdots&\vdots&\vdots&\vdots\\
\varphi_{n}&\lambda_{n}\phi_{n}&\ldots&\lambda_{n}^{n-3}\phi_{n}&\lambda_{n}^{n-2}\varphi_{n}&-\lambda_{n}^{n}\phi_{n}\\
\end{vmatrix},
\end{equation}
\begin{equation*}
\Omega_{12}=\begin{vmatrix}
\varphi_{1}&\lambda_{1}\phi_{1}&\ldots&\lambda_{1}^{n-3}\phi_{1}&\lambda_{1}^{n-2}\varphi_{1}&\lambda_{1}^{n-1}\varphi_{1}\\
\varphi_{2}&\lambda_{2}\phi_{2}&\ldots&\lambda_{2}^{n-3}\phi_{2}&\lambda_{2}^{n-2}\varphi_{2}&\lambda_{2}^{n-1}\varphi_{2}\\
\vdots&\vdots&\vdots&\vdots&\vdots&\vdots\\
\varphi_{n}&\lambda_{n}\phi_{n}&\ldots&\lambda_{n}^{n-3}\phi_{n}&\lambda_{n}^{n-2}\varphi_{n}&\lambda_{n}^{n-1}\varphi_{n}\\
\end{vmatrix},
\end{equation*}

\begin{equation*}
\Omega_{21}=\begin{vmatrix}
\phi_{1}&\lambda_{1}\varphi_{1}&\ldots&\lambda_{1}^{n-3}\varphi_{1}&\lambda_{1}^{n-2}\phi_{1}&-\lambda_{1}^{n}\varphi_{1}\\
\phi_{2}&\lambda_{2}\varphi_{2}&\ldots&\lambda_{2}^{n-3}\varphi_{2}&\lambda_{2}^{n-2}\phi_{2}&-\lambda_{2}^{n}\varphi_{2}\\
\vdots&\vdots&\vdots&\vdots&\vdots&\vdots\\
\phi_{n}&\lambda_{n}\varphi_{n}&\ldots&\lambda_{n}^{n-3}\varphi_{n}&\lambda_{n}^{n-2}\phi_{n}&-\lambda_{n}^{n}\varphi_{n}\\
\end{vmatrix},
\end{equation*}
\begin{equation*}
\Omega_{22}=\begin{vmatrix}
\phi_{1}&\lambda_{1}\varphi_{1}&\ldots&\lambda_{1}^{n-3}\varphi_{1}&\lambda_{1}^{n-2}\phi_{1}&\lambda_{1}^{n-1}\phi_{1}\\
\phi_{2}&\lambda_{2}\varphi_{2}&\ldots&\lambda_{2}^{n-3}\varphi_{2}&\lambda_{2}^{n-2}\phi_{2}&\lambda_{2}^{n-1}\phi_{2}\\
\vdots&\vdots&\vdots&\vdots&\vdots&\vdots\\
\phi_{n}&\lambda_{n}\varphi_{n}&\ldots&\lambda_{n}^{n-3}\varphi_{n}&\lambda_{n}^{n-2}\phi_{n}&\lambda_{n}^{n-1}\phi_{n}\\
\end{vmatrix}.
\end{equation*}

We are now in a position to consider the reduction of the DT of the GI system so that
$q^{[n]}=-(r^{[n]})^*$, then the DT of the GI is given. Under the reduction condition $q=-r^*$,
the eigenfunction $\psi_k=\left( \begin{array}{c}
\phi_k\\
\varphi_k
\end{array} \right)$ associated with eigenvalue $\lambda_k$ has following properties,
\begin{description}
\item[(i)] $\phi_{k}^{\ast}=\varphi_{k}$, $\lambda_{k}=-{\lambda_{k}}^{\ast}$;
\item[(ii)] ${\phi_{k}}^{\ast}=\varphi_{l}, \ -{\varphi_{k}}^{\ast}=\phi_{l}$,
${\lambda_{k}}^{\ast}=\lambda_{l}$, where $k\neq l$.
\end{description}
Notice that the denominator $W_n$ of $q^{[n]}$ is a non-zero complex function under
reduction condition when $\lambda_i\not=0$, so the new solution $q^{[n]}$ is non-singular.
We shall explain it clearly in following Case 1. For the one-fold DT $T_1$, set
\begin{equation}\label{onefoldredu}
 \lambda_1=i\beta_1\text{(a pure imaginary),\quad and its eigenfunction \quad}
\psi_1=\left( \begin{array}{c}
\phi_1\\
\phi_1^*
\end{array} \right),
\end{equation}
then $T_1$ in theorem 1 is the DT of the GI.
We note that $q^{[1]}=-(r^{[1]})^*$ holds  with the help of eq.(\ref{sTT}), $q=-r^*$ and this special
choice of $\psi_1$. This is an essential distinctness of DT between GI and NLS, because one-fold
transformation of AKNS can not preserve the reduction condition to the NLS. Furthermore, for the
two-fold DT, according to above property (ii), set
\begin{equation}\label{twofoldredu1}
\lambda_2=\lambda_1^*\text{\quad and its eigenfunction }
\psi_2=\left( \begin{array}{c}
-\varphi_1^*\\
\phi_1^*
\end{array} \right),
\psi_1=\left( \begin{array}{c}
\phi_1\\
\varphi_1
\end{array} \right) \text{associated with eigenvalue\quad } \lambda_1,
\end{equation}
then $q^{[2]}=-(r^{[2]})^*$ can be verified from eq. (\ref{ntt4})
and $T_2$ given by eq.(\ref{fss1}) is the DT of the GI.
Of course,  in order to get $q^{[2]}=-(r^{[2]})^*$ so that $T_2$ becomes also
the DT of the DNLS, we  can also set
\begin{equation}\label{twofoldredu2}
\lambda_l=i\beta_l \text{(a pure imaginary) and its eigenfunction\ }
\psi_l=\left( \begin{array}{c}
\varphi_l^*\\
\phi_l^*
\end{array} \right),l=1,2.
\end{equation}
There are many choices to guarantee  $q^{[n]}=-(r^{[n]})^*$ for the n-fold DTs when $n> 2$. For example,
setting $n=2k$ and $l=1,3,\dots,2k-1$,then choosing following $k$ distinct
eigenvalues and eigenfunctions in n-fold DTs:
\begin{equation}\label{2nfoldredu}
 \lambda_l \leftrightarrow \psi_l=\left( \begin{array}{c}
\phi_l\\
\varphi_l\end{array} \right)
, \text{and} \lambda_{2l}= \lambda_{2l-1}^*,\leftrightarrow
\psi_{2l}=\left( \begin{array}{c}
-\varphi_{2l-1}^*\\
\phi_{2l-1}^*
\end{array} \right)
\end{equation}
so that $q^{[2k]}=-(r^{[2k]})^*$ in eq.(\ref{ntt4}). Then $T_{2k}$
with these paired-eigenvalue $\lambda_i$ and paired-eigenfunctions
$\psi_i(i=1,3,\dots,2k-1)$ is reduced to the (2k)-fold DT of the GI.
Similarly, $T_{2k+1}$ in eq.(\ref{fss2}) can also be reduced to the
(2k+1)-fold DT of the GI by choosing one pure imaginary
$\lambda_{2k+1}=i\beta_{2k+1}(\text{pure imaginary})$ and  $k$
paired-eigenvalues $\lambda_{2l}= \lambda_{2l-1}^*(l=1,2,\cdots,k)$
with corresponding eigenfunctions according to properties (i) and
(ii).
\section{breather solution and rogue wave}

Similar to case of the DNLSI\cite{kenji1,steduel,xuhe}, it is not difficult to get new solutions
of the GI equation from zero ``seed'' and nonzero `` seed'' by one-fold DT. We shall focus directly
on BA solutions and RW from a periodic ``seed" by 2-fold and 4-fold DT.

Set $a$ and  $c$ be two complex constants,
then $ q = ce^{{\rm i}[ax-(a^{2}+c^{2}a-\frac{c^{4}}{2})t]}$ is a periodic solution of the GI equation, which will be used as
a ``seed'' solution of the DT. Substituting $q = ce^{{\rm i}[ax-(a^{2}+c^{2}a-\frac{c^{4}}{2})t]}$ into the spectral problem eq.(\ref{sys11})
 and eq.(\ref{sys22}), and using the method of separation of variables and the superposition
principle, the eigenfunction $\psi_k$ associated with $\lambda_k$ is given by
\begin{eqnarray}\label{eigenfunfornonzeroseed}
\left(\mbox{\hspace{-0.2cm}} \begin{array}{c}
 \phi_{k}(x,t,\lambda_{k})\\
 \varphi_{k}(x,t,\lambda_{k})\\
\end{array}\mbox{\hspace{-0.2cm}}\right)\mbox{\hspace{-0.2cm}}=\mbox{\hspace{-0.2cm}}\left(\mbox{\hspace{-0.2cm}}\begin{array}{c}
 \varpi1(x,t,\lambda_{k})[1,k]+\varpi2(x,t,\lambda_{k})[1,k]-\varpi1^{\ast}(x,t,{\lambda_{k}^{\ast})}[2,k]-\varpi2^{\ast}(x,t,{\lambda_{k}^{\ast})}[2,k]\\
 \varpi1(x,t,\lambda_{k})[2,k]+\varpi2(x,t,\lambda_{k})[2,k]+\varpi1^{\ast}(x,t,{\lambda_{k}^{\ast})}[1,k]+\varpi2^{\ast}(x,t,{\lambda_{k}^{\ast})}[1,k]\\
\end{array}\mbox{\hspace{-0.2cm}}\right).
\end{eqnarray}
Here
\begin{eqnarray*}
\left(\mbox{\hspace{-0.2cm}}\begin{array}{c}
 \varpi1(x,t,\lambda_{k})[1,k]\\
 \varpi1(x,t,\lambda_{k})[2,k]\\
\end{array}\mbox{\hspace{-0.2cm}}\right)\mbox{\hspace{-0.2cm}}=\mbox{\hspace{-0.2cm}}\left(\mbox{\hspace{-0.2cm}}\begin{array}{c}
\exp(\frac{-\sqrt{s}(ta-x-2t{\lambda_{k}}^2)}{2}+\frac{{\rm i}[ax-(a^{2}+c^{2}a-\frac{c^{4}}{2})t]}{2}) \\
\frac{({\rm i}a-{\rm i}c^{2}+2{\rm i}{\lambda_{k}}^2+\sqrt{s})}{2{\lambda_{k}}c}\exp(\frac{-\sqrt{s}(ta-x-2t{\lambda_{k}}^2)}{2}-\frac{{\rm i}[ax-(a^{2}+c^{2}a-\frac{c^{4}}{2})t]}{2}) \\\\
\end{array}\mbox{\hspace{-0.2cm}}\right),\\
\end{eqnarray*}
\begin{eqnarray*}
\left(\mbox{\hspace{-0.2cm}}\begin{array}{c}
 \varpi2(x,t,\lambda_{k})[1,k]\\
 \varpi2(x,t,\lambda_{k})[2,k]\\
\end{array}\mbox{\hspace{-0.2cm}} \right)\mbox{\hspace{-0.2cm}}=\mbox{\hspace{-0.3cm}}
\left(\mbox{\hspace{-0.3cm}}\begin{array}{c}
\exp(\frac{\sqrt{s}(ta-x-2t{\lambda_{k}}^2)}{2}+\frac{{\rm i}[ax-(a^{2}+c^{2}a-\frac{c^{4}}{2})t]}{2}) \\
\frac{({\rm i}a-{\rm i}c^{2}+2{\rm i}{\lambda_{k}}^2-\sqrt{s})}{2{\lambda_{k}}c}\exp(\frac{\sqrt{s}(ta-x-2t{\lambda_{k}}^2)}{2}-\frac{{\rm i}[ax-(a^{2}+c^{2}a-\frac{c^{4}}{2})t]}{2}) \\
\end{array}\mbox{\hspace{-0.3cm}}\right)\mbox{\hspace{-0.1cm}},\\
\end{eqnarray*}
\begin{eqnarray*}
\varpi1(x,t,\lambda_{k})=
\left( \begin{array}{c}
 \varpi1(x,t,\lambda_{k})[1,k]\\
 \varpi1(x,t,\lambda_{k})[2,k]\\
\end{array} \right),~~~~~
\varpi2(x,t,\lambda_{k})=
\left( \begin{array}{c}
 \varpi2(x,t,\lambda_{k})[1,k]\\
 \varpi2(x,t,\lambda_{k})[2,k]\\
\end{array} \right),
\end{eqnarray*}
\begin{eqnarray}
&s=-c^{4}+2c^{2}a-a^{2}-4a{\lambda_{k}}^{2}-4{\lambda_{k}}^{4}. \nonumber
\end{eqnarray}
Note that
$\varpi1(x,t,\lambda_{k})$ and $\varpi2(x,t,\lambda_{k})$ are two different solutions of
the spectral problem eq.(\ref{sys11}) and eq.(\ref{sys22}),
but we  can only get the  trivial solutions through DT of the GI by setting eigenfunction $\psi_k$ be one of them.

  Using eigenfunctions $\psi_k$ in eq.(\ref{eigenfunfornonzeroseed}), following several examples
give a transparent explanation of the DT of the GI equation.

\noindent Case 1.($N = 2$).Under the choice in eq.(\ref{twofoldredu1}) with one paired eigenvalue
$\lambda_{1}=\alpha_{1}+i\beta_{1}$ and $\lambda_{2}=\alpha_{1}-i\beta_{1}$,
the two-fold DT eq.(\ref{ntt4}) of the GI equation implies  a solution
\begin{equation}\label{q2j2}
 q^{[2]}=q+2i\dfrac{({\lambda_{2}}^{2}-{\lambda_{1}}^{2})\phi_{1}{\varphi_{1}}^{\ast}}{\lambda_{1}\varphi_{1}{\varphi_{1}}^{\ast}+\lambda_{2}\phi_{1}{\phi_{1}}^{\ast}},
\end{equation}
with $\phi_{1}$ and $\varphi_{1}$ given by eq.(\ref{eigenfunfornonzeroseed}).
Note the denominator in eq.(\ref{q2j2}) is non-zero except $\lambda_1=0$, because the
imaginary part of denominator will be zero if and only if  $\alpha_1=\beta_1=0$. Two concrete examples of eq.(\ref{q2j2})
are given below.\\
\noindent (a)For simplicity, let $a=-2{\alpha_{1}}^{2}+2{\beta_{1}}^{2}$ so that
$\text{Im}(-c^{4}+2c^{2}a-a^{2}-4a{\lambda_{1}}^{2}-4{\lambda_{1}}^{4})=0$, then
\begin{eqnarray}\label{q2j3}
q^{[2]}=\exp(i(2xS+(-4S^{2}-2c^{2}S+\dfrac{1}{2}c^{4})t))(c+\dfrac{u1}{u2}),
\end{eqnarray}
\begin{eqnarray*}
&&u1=\lefteqn{4\alpha_{1}\beta_{1}((c-2\beta_{1})^{2}(c^2+4\alpha_{1}^2)-K1^2)\cosh(f1)-16i{\alpha_{1}}^{2}\beta_{1}(c-2\beta_{1})K1\sin(f2)
{}}\nonumber\\&&\mbox{\hspace{1cm}}+4\alpha_{1}\beta_{1}((c-2\beta_{1})^{2}(c^2+4\alpha_{1}^2)+K1^2)\cos(f2)-8ic\alpha_{1}\beta_{1}(c-2\beta_{1})K1\sinh(f1),\\
&&u2=\lefteqn{\alpha_{1}((c-2\beta_{1})^{2}(c^2+4\alpha_{1}^2)+K1^2)\cosh(f1)-2ic\beta_{1}(c-2\beta_{1})K1\sin(f2)
{}}\nonumber\\
&&\mbox{\hspace{1cm}}+\alpha_{1}((c-2\beta_{1})^{2}(c^2+4\alpha_{1}^2)-K1^2)\cos(f2)+4i\alpha_{1}\beta_{1}(c-2\beta_{1})K1\sinh(f1),\\
&&S={\beta_{1}}^{2}-{\alpha_{1}}^{2},\;\; f1=K1(4({\alpha_{1}}^{2}-{\beta_{1}}^{2})t+x),\;\;f2=4K1\alpha_{1}\beta_{1}t,\\
&&K1=\sqrt{16{\alpha_{1}}^{2}{\beta_{1}}^{2}-4c^{2}{\alpha_{1}}^{2}+4c^{2}{\beta_{1}}^{2}-c^{4}}.
\end{eqnarray*}
By letting $ x \rightarrow {\infty}, \ t \rightarrow {\infty}$, so $|q^{[2]}|^{2} \rightarrow c^2$,
the trajectory of this solution is defined explicitly by
\begin{equation}\label{q1j4}
x=-4{\alpha_{1}}^{2}t+4{\beta_{1}}^{2}t
 \end{equation}
from $f_1=0$
if $K1^2>0$, and by
\begin{equation}
t=0
\end{equation}
from $f_2=0$ if $K1^2<0$. According to eq.(\ref{q2j3}), we can get the Ma breathers\cite{MaB}(time periodic
breather solution) and the Akhmediev breathers\cite{Akhmediev1,Akhmediev2} (space periodic breather solution)
solution. In general, the solution in eq.(\ref{q2j3}) evolves periodically along the  straight line
with a certain angle of $x$ axis and $t$ axis. The dynamical evolution of
$|q^{[2]}|^2$ in eq.(\ref{q2j3}) for different parameters are plotted in
 Figure 1, Figure 2 and Figure 3, which give a visual verification of the three cases of
 trajectories. As we have done for the DNLSI\cite{xuhe}, the rogue wave of the GI equation
 will be given from its BA solutions by a limit procedure in the following.
 By letting $ c \rightarrow 2\beta_{1}$ in
 $(\ref{q2j3})$ with $\text{Im}(-c^{4}+2c^{2}a-a^{2}-4a{\lambda_{1}}^{2}-4{\lambda_{1}}^{4})=0$,
it becomes rogue wave
\begin{equation}\label{rationaljie}
{q_{rogue~wave}}^{[2]}=\dfrac{k1}{k2}+k3.
\end{equation}
Here
\begin{eqnarray*}
&&k1=\lefteqn{-4\beta_1(16 i\alpha_1 \beta_1^{2}(\beta_1^{2}+\alpha_1^{2}) t+\beta_1
i+\alpha_1-4\beta_1(\beta_1^{2}+\alpha_1^{2}) x-16\beta_1(\alpha_1^{4}-
\beta_1^{4}) t){}}\nonumber\\&&{}
\mbox{\hspace{0.8cm}}\times(16 i \alpha_1\beta_1^{2}(\beta_1^{2}+\alpha_1^{2}) t-i
\beta_1+\alpha_1+4\beta_1(\beta_1^{2}+\alpha_1^{2}) x+16\beta_1(\alpha_1^{4}
-\beta_1^{4}) t)\nonumber\\&&{}\mbox{\hspace{0.8cm}}\times\exp(-2 i ((\alpha_1^2-\beta_1^{2}) x+2 (\alpha_1^{4}-4 \alpha_1^{2} \beta_1^{2}+\beta_1^{4})t)),\\
&&k2=\lefteqn{-16\beta_1^{2}(\beta_1^{2}+\alpha_1^{2})(16(\alpha_1^{6}+
\beta_1^{6})t^2-8(-\alpha_1^{4}+\beta_1^{4}) x t+(\beta_1^{2}+\alpha_1^{2})x^2)
-\beta_1^2-\alpha_1^{2}{}}\nonumber\\&&{}
\mbox{\hspace{0.8cm}}+32i\beta_1^{2}(\alpha_1^{2}-2\beta_1^{2})(\beta_1^{2}+\alpha_1^{2})t+8i\beta_1^{2}(\beta_1^{2}+\alpha_1^{2}) x,\\
&&k3=2 \beta_{1} \exp(-2 i ((\alpha_1^2-\beta_1^{2}) x+2 (\alpha_1^{4}-4 \alpha_1^{2} \beta_1^{2}+\beta_1^{4})t)).
\end{eqnarray*}
By letting $ x \rightarrow {\infty}, \ t \rightarrow {\infty}$,
 so $|{q_{rogue~wave}}^{[2]}|^{2}\rightarrow 4\beta_{1}^2$.
The rouge wave reaches its  maximum amplitude $|{q_{rogue~wave}}^{[2]}|^{2}=36\beta_{1}^{2}$
at $t = 0$ and $x=0$  and  minimum amplitude  $|{q_{rogue~wave}}^{[2]}|^{2}=0$ at
  $t = \pm\dfrac{3}{16\sqrt{3(4\alpha_{1}^{2}+\beta_{1}^{2})}\beta_{1}(\alpha_{1}^{2}
  +\beta_{1}^{2})}$ and $x=\mp\dfrac{9\alpha_{1}^{2}}{4\sqrt{3(4\alpha_{1}^{2}+\beta_{1}^{2})}
  \beta_{1}(\alpha_{1}^{2}+\beta_{1}^{2})}$. Figure 4 and Figure 5 of
$|{q_{rogue~wave}}^{[2]}|^2$ show the  the maximum value of $|{q_{rogue~wave}}^{[2]}|^2$
 is nine times of its asymptotic value and there exists only a very small domain on whole
 (x-t) plane of non-constant value. Therefore ${q_{rogue~wave}}^{[2]}$ is one kind of typical
 rogue wave.

\noindent(b)When $a=\dfrac{c^2}{2}$, it is not difficult  from eq.(\ref{eigenfunfornonzeroseed}) to
find that  there are two sets of collinear eigenfunctions,
\begin{equation}\label{xiangguan1}
\left( \begin{array}{c}
 \varpi1(x,t,\lambda_{k})[1,k]\\
 \varpi1(x,t,\lambda_{k})[2,k]\\
\end{array} \right) \text{and} \ \
\left( \begin{array}{c}
-\varpi2^{\ast}(x,t,{\lambda_{k}^{\ast})}[2,k]\\
\varpi2^{\ast}(x,t,{\lambda_{k}^{\ast})}[1,k]\\
\end{array}\right),\\
\end{equation}
\begin{equation}\label{xiangguan2}
\left( \begin{array}{c}
 \varpi2(x,t,\lambda_{k})[1,k]\\
 \varpi2(x,t,\lambda_{k})[2,k]\\
\end{array} \right)\text{and} \ \
\left( \begin{array}{c}
-\varpi1^{\ast}(x,t,{\lambda_{k}^{\ast})}[2,k]\\
\varpi1^{\ast}(x,t,{\lambda_{k}^{\ast})}[1,k]\\
\end{array} \right).
\end{equation}
Therefore,  the eigenfunction $\psi_k$ associated with $\lambda_k$
 for this case is given by
\begin{eqnarray}\label{eigenfunfornonzeroseed1}
\left(\mbox{\hspace{-0.2cm}} \begin{array}{c}
 \phi_{k}(x,t,\lambda_{k})\\
 \varphi_{k}(x,t,\lambda_{k})\\
\end{array}\mbox{\hspace{-0.2cm}}\right)\mbox{\hspace{-0.2cm}}=\mbox{\hspace{-0.2cm}}\left(\mbox{\hspace{-0.2cm}}\begin{array}{c}
 \varpi1(x,t,\lambda_{k})[1,k]-\varpi1^{\ast}(x,t,{\lambda_{k}^{\ast})}[2,k]\\
 \varpi1(x,t,\lambda_{k})[2,k]+\varpi1^{\ast}(x,t,{\lambda_{k}^{\ast})}[1,k]\\
\end{array}\mbox{\hspace{-0.2cm}}\right).
\end{eqnarray}
Here
\begin{eqnarray*}
\left(\mbox{\hspace{-0.2cm}}\begin{array}{c}
 \varpi1(x,t,\lambda_{k})[1,k]\\
 \varpi1(x,t,\lambda_{k})[2,k]\\
\end{array}\mbox{\hspace{-0.2cm}}\right)\mbox{\hspace{-0.2cm}}=\mbox{\hspace{-0.2cm}}\left(\mbox{\hspace{-0.2cm}}\begin{array}{c}
\exp(i({\lambda_{k}}^{2}x+2{\lambda_{k}}^{4}t+\dfrac{1}{2}c^2 x-\dfrac{1}{4} c^{4} t)) \\
\dfrac{2 i \lambda_{k}}{c }\exp(i({\lambda_{k}}^{2}x+2{\lambda_{k}}^{4}t)) \\
\end{array}\mbox{\hspace{-0.2cm}}\right).
\end{eqnarray*}
Under the choice in eq.(\ref{twofoldredu1}) with
$\lambda_{1}=\alpha_{1}+i\beta_{1},\lambda_{2}=\alpha_{1}-i\beta_{1}$,
and the $\psi_1$ given by eq.(\ref{eigenfunfornonzeroseed1}), the
solution $q^{[2]}$ is given simply from eq. (\ref{ntt4}). Figure 6
is plotted for $|q^{[2]}|^2$, which shows the  periodical
evolution along a straight line on $(x-t)$ plane.
\\
Case 2.($N = 4$). According to the choice in eq.(\ref{2nfoldredu}) with two distinct  eigenvalues
$\lambda_1=\alpha_1+i\beta_1, \lambda_3=\alpha_3+i\beta_3$, substituting $\psi_1$ and $\psi_3$
defined by eq. (\ref{eigenfunfornonzeroseed}) into eq.(\ref{ntt4}), then the new solution
$q^{[4]}$ generated by 4-fold DT is given. Its analytical expression is omitted because it is
very complicated. But $|q^{[4]}|^2$ are plotted in Figure 7 and 8 to show the
dynamical evolution on $(x-t)$ plane:
(a) Let $a=-2{\alpha_{i}}^{2}+2{\beta_{i}}^{2},i=1,3$, so
that $\text{Im}(-c^{4}+2c^{2}a-a^{2}-4a{\lambda_{i}}^{2}-4{\lambda_{i}}^{4})=0$, then Figure 7 shows
intuitively that two breathers may have parallel trajectories;(b) Two breathers
have an elastic collision so that they can preserve their profiles after interaction, which
is verified in Figure 8.
\section{Conclusions and discussions}

 The determinant representation of DT for the GI system has been constructed in Thm1 and Thm2.
 By taking paired eigenvalue $\lambda_2=\lambda_1^*$ and associated eigenfunctions in two-fold DT,
 then reduction condition $q^{[2]}=-(r^{[2]})^*$ is preserved and the BA solution and RW of the
 GI equation are obtained analytically from a periodic ``seed''. In particular, the kernel of the
 one-fold DT of the GI system is one-dimensional, which is  true for GI equation when the eigenvalue
 is pure imaginary. Furthermore, $q^{[n]}$ in eq.(\ref{ntt4}) is also a solution of the GI equation
 if the eigenvalues and eigenfunctions in n-fold DT are chosen according to reduction conditions
 given at the last paragraph of the section 2. Eight figures are plotted for the BA and RW of the
 GI equation to show their dynamical evolution on (x-t) plane.

   For the one-fold DT, the DNLSI and GI equation have same form as eq.(\ref{TT}). The
 coefficient  of the first order $\lambda$ is a constant matrix and the other is
 a matrix of function of ($x,t$) for GI, but it is in a opposite case for the DNLSI(see eq.(17) of
 reference\cite{xuhe}). This implies that $q^{[2]}$ eq.(\ref{q2j2}) of the GI is simpler than eq.(52)
 in reference\cite{xuhe} of the DNLSI. Furthermore, the RW of the GI is also simpler than the
 one (eq.(56) of reference\cite{xuhe}) of the DNLSI. Although the DT of the GI has also been given
 before in reference\cite{fan1,fan2}, there are two new points in present paper:
 1) the one-fold DT has a one-dimensional kernel; 2) the BA solution and RW  are new,
 which are constructed from a non-zero ``seed", because Fan has just got solitons from zero
 ``seed" by DT. Our results show several interesting differences between DNLSI and GI, which
supports again that  each of them deserves studying separately. Moreover, the appearance  of
BA and RW for GI  also reminds us possibly that GI will be an interesting physical model in optical
system under consideration of higher order nonlinear effects.

{\bf Acknowledgments}
{\noindent \small  This work is supported by the NSF of China under Grant No.10971109 and K.C.Wong Magna Fund
in Ningbo University. Jingsong He is also supported by Program for NCET under Grant No.NCET-08-0515.Shuwei Xu is also supported by the Scientific
Research Foundation of Graduate School of Ningbo University. We thank
Prof. Yishen Li(USTC,Hefei, China) for his useful suggestions on the rogue wave.}

\begin{figure}[htbp]     \centering \mbox{}\hspace{0cm}
      \includegraphics[width=.3\textwidth]{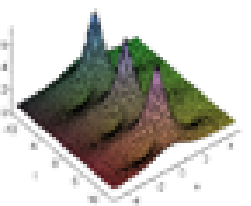}
      \mbox{}\vspace{0.5cm}\caption{
The
dynamical evolution of $|q^{[2]}|^2$({\bf time periodic} breather)
in eq.(\ref{q2j3}) on ($x-t$)
plane with specific parameters
$\alpha_{1}=\beta_{1},\beta_{1}=0.5,c=0.75$. The trajectory is a line $x=0$. }
\end{figure}

\begin{figure}[htbp]     \centering \mbox{}\hspace{0cm}
      \includegraphics[width=.3\textwidth]{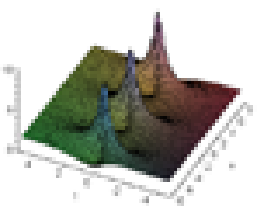}
      \mbox{}\vspace{0.5cm}\caption{
The
dynamical evolution of $|q^{[2]}|^2$({\bf space periodic} breather) in eq.(\ref{q2j3})
on ($x-t$) plane with specific parameters
$\alpha_{1}=\beta_{1},\beta_{1}=0.5,c=1.2$. The trajectory is a line $t=0$. }
\end{figure}

\begin{figure}[htbp]     \centering \mbox{}\hspace{0cm}
      \includegraphics[width=.3\textwidth]{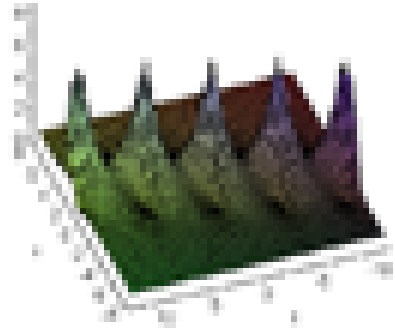}
      \mbox{}\vspace{0.5cm}\caption{
The
dynamical evolution of solution $|q^{[2]}|^2$ in eq.(\ref{q2j3}) for case 1 (a).
It evolves periodically along a straight line with
certain angle of $x$ axis and $t$ axis under specific parameters
$\alpha_{1}=0.5,\beta_{1}=0.55,c=0.75$. }
\end{figure}

\begin{figure}[htbp]     \centering \mbox{}\hspace{0cm}
      \includegraphics[width=.3\textwidth]{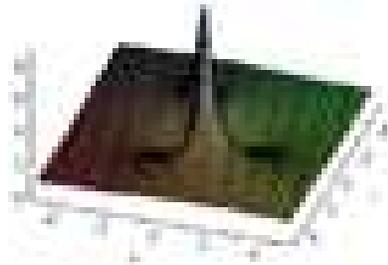}
      \mbox{}\vspace{0.5cm}\caption{
The
dynamical evolution of $|q^{[2]}_{rogue~wave}|^{2}$ given by eq.(\ref{rationaljie})  on ($x-t$) plane
 with specific parameters
$\alpha_{1}=\dfrac{1}{2},\beta_{1}=\dfrac{1}{2}$.By letting $ x \rightarrow {\infty}, \ t \rightarrow {\infty}$, so $|q^{[2]}_{rogue~wave}|^{2}\rightarrow 1$.
The maximum amplitude of $|q^{[2]}_{rogue~wave}|^{2}$ occurs at $t = 0$ and $x=0$ and is equal to 9,and the minimum amplitude of $|q^{[2]}_{rogue~wave}|^{2}$ occurs at $t = \pm\dfrac{\sqrt{15}}{10}$ and $x=\mp\dfrac{3\sqrt{15}}{10}$ and is equal to $0$. }
\end{figure}

\begin{figure}[htbp]     \centering \mbox{}\hspace{0cm}
      \includegraphics[width=.3\textwidth]{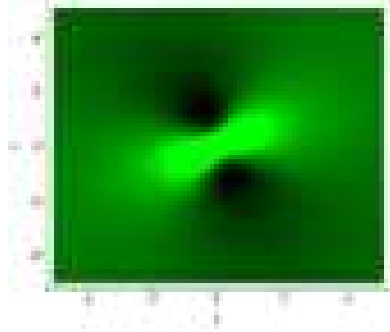}
      \mbox{}\vspace{0.5cm}\caption{
Contour plot of the wave amplitudes of
$|q^{[2]}_{rogue~wave}|^{2}$ in the ($x-t$) plane is given by eq.(\ref{rationaljie}) for $\alpha_{1}=\dfrac{1}{2},\beta_{1}=\dfrac{1}{2}$. }
\end{figure}

\begin{figure}[htbp]     \centering \mbox{}\hspace{0cm}
      \includegraphics[width=.3\textwidth]{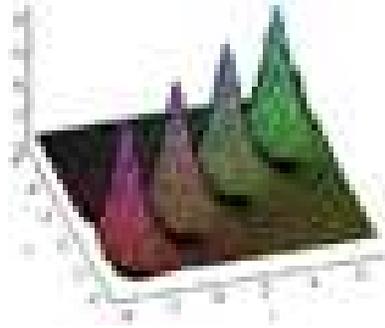}
      \mbox{}\vspace{0.5cm}\caption{
The
dynamical evolution of $|q^{[2]}|^2$ in case 1(b) on ($x-t$) plane with specific parameters
$\alpha_{1}=0.65,\beta_{1}=0.5,c=0.95$. It evolves periodically along a straight line on $(x-t)$
plane. }
\end{figure}

\begin{figure}[htbp]     \centering \mbox{}\hspace{0cm}
      \includegraphics[width=.3\textwidth]{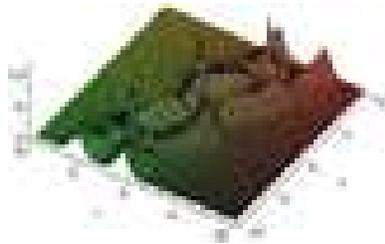}
      \mbox{}\vspace{0.5cm}\caption{
The
dynamical evolution of  periodic breather
solution given by case 2(a) on ($x-t$) plane
 with specific parameters
$\alpha_{1}=0.5,\beta_{1}=0.4,c=0.35,\alpha_{3}=0.6,\beta_{3}=\dfrac{3\sqrt{3}}{10}$.
This picture shows two breathers may parallelly propagate on ($x-t$) plane. }
\end{figure}

\begin{figure}[htbp]     \centering \mbox{}\hspace{0cm}
      \includegraphics[width=.3\textwidth]{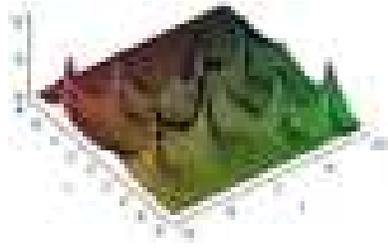}
      \mbox{}\vspace{0.5cm}\caption{
The
dynamical evolution of periodic breather
solution given by case 2(b) on ($x-t$) plane  with specific parameters
$a=\dfrac{c^2}{2},\alpha_{1}=-0.5,\beta_{1}=0.5,\alpha_{3}=0.6,\beta_{3}=0.5,c=0.75$.
This picture shows the elastic interaction of the two breathers. }
\end{figure}
\end{document}